\documentclass[final]{aipproc}

\layoutstyle{6x9}

\begin{document}

\title{Genuine Four Tangle for Four Qubit States}


\author{S. Shelly Sharma}{
  address={Depto. de Fisica, Universidade Estadual de Londrina, Londrina 86051-990, PR Brazil}
}

\author{N. K. Sharma}{
  address={Depto. de Matematica, Universidade Estadual de Londrina, Londrina 86051-990, PR Brazil}
}

\begin{abstract}
 We report a four qubit polynomial invariant that quantifies genuine four-body correlations. 
The four qubit invariants are obtained from transformation properties of three qubit invariants
 under a local unitary on the fourth qubit. 
\end{abstract}

\maketitle

Two multipartite pure states are equivalent under stochastic local
operations and classical communication (SLOCC) \cite{vers02} if one
can be obtained from the other with some probability using SLOCC. Attempts 
\cite{lama07,li09} to classify four-qubit pure states under SLOCC,
have revealed that several entanglement classes contain a continuous range
of strictly nonequivalent states, although with similar structure. In view
of this, we proposed classification criteria \cite{shar12} based on nature
of multiqubit correlations in N-qubit pure states. In this article, we
examine the three qubit invariants of four qubit states and derive higher
degree invariants to quantify four and three-way correlations.

For a two qubit state, negative eigenvalue of
partially transposed state operator is the invariant that distinguishes between a
separable and an entangled state. In three qubit state space, 
two qubit subspace (for a selected pair of qubits) is
characterized by a pair of two qubit invariants, while new two qubit
invariants arise due to three body correlations in the composite space. The
most important three qubit polynomial invariant is a degree four combination
of two qubit invariants. The entanglement monotone constructed from this is
Wootter's three tangle \cite{coff00}.  Four qubit states sit in the space $C^{2}\otimes
C^{2}\otimes C^{2}\otimes C^{2}$ with three qubit subspaces for each set of
three qubits. If there were no four body correlations, then three tangles
should determine the entanglement of a four qubit state. When four body
correlations are present, additional three qubit invariants that depend on
four way negativity fonts \cite{shar102} exist. Three
qubit invariants, for a given set of three qubits, constitute a five dimensional space. In this article, we
obtain four qubit invariants using transformation properties of three qubit
invariants under a local unitary applied to the fourth qubit. One can
continue the process to a higher number of qubits. 

\section{Five Three tangles}
Two qubit unitary invariants for pair of qubits $A_{1}A_{2}$ in the most
general four qubit state%
\begin{equation}
\left\vert \Psi ^{A_{1}A_{2}A_{3}A_{4}}\right\rangle
=\sum_{i_{1}i_{2}i_{3}i_{4}}a_{i_{1}i_{2}i_{3}i_{4}}\left\vert
i_{1}i_{2}i_{3}i_{4}\right\rangle ;\quad \left( i_{m}=0,1\right) ,
\end{equation}%
are $D_{\left( A_{3}\right) _{i_{3}}\left( A_{4}\right) _{i_{4}}}^{00}$, $%
D_{\left( A_{3}\right) _{i_{3}}}^{00i_{4}}-D_{\left( A_{3}\right)
_{i_{3}}}^{01i_{4}}$, $D_{\left( A_{4}\right) _{i_{4}}}^{00i_{3}}-D_{\left(
A_{4}\right) _{i_{4}}}^{01i_{3}}$, $D^{0000}-D^{0100}$, $D^{0001}-D^{0101}$,
where%
\begin{equation}
D_{\left( A_{3}\right) _{i_{3}}\left( A_{4}\right) _{i_{4}}}^{00}=\det \left[
\begin{array}{cc}
a_{00i_{3}i_{4}} & a_{01i_{3}i_{4}} \\ 
a_{10i_{3}i_{4}} & a_{11i_{3}i_{4}}%
\end{array}%
\right], D^{0i_{2}0i_{4}}=\det \left[ 
\begin{array}{cc}
a_{0i_{2}0i_{4}} & a_{0,i_{2}+1,1,i_{4}+1} \\ 
a_{1i_{2}0i_{4}} & a_{1,i_{2}+1,1,i_{4}+1}%
\end{array}%
\right] ,
\end{equation}%
\begin{equation}
D_{\left( A_{3}\right) _{i_{3}}}^{0i_{2}0}=\det \left[ 
\begin{array}{cc}
a_{0i_{2}i_{3}0} & a_{0i_{2}+1i_{3}1} \\ 
a_{1i_{2}i_{3}0} & a_{1i_{2}+1i_{3}1}%
\end{array}%
\right] , D_{\left( A_{4}\right) _{i_{4}}}^{0i_{2}0}=\det \left[ 
\begin{array}{cc}
a_{0i_{2}0i_{4}} & a_{0i_{2}+1,1i_{4}} \\ 
a_{1i_{2}0i_{4}} & a_{1i_{2}+1,1i_{4}}%
\end{array}%
\right] .
\end{equation}%

For qubits $A_{1}A_{2}A_{3}$ in $\left\vert \Psi
^{A_{1}A_{2}A_{3}A_{4}}\right\rangle $, three qubit invariants
\begin{equation}
\left( I_{3}^{A_{1}A_{2}A_{3}}\right) _{\left( A_{4}\right) _{i_{4}}}=\left(
D_{\left( A_{4}\right) _{i_{4}}}^{000}-D_{\left( A_{4}\right)
_{i_{4}}}^{010}\right) ^{2}-4D_{\left( A_{3}\right) _{0}\left( A_{4}\right)
_{i_{4}}}^{00}D_{\left( A_{3}\right) _{1}\left( A_{4}\right)
_{i_{4}}}^{00};\quad i_{4}=0,1.
\end{equation}%
quantify GHZ state like three-way correlations in three qubit state space.
We examine the action of U$^{A_{4}}=\frac{1}{\sqrt{1+\left\vert y\right\vert ^{2}}}\left[
\begin{array}{cc}
1 & -y^{\ast } \\ 
y & 1%
\end{array}%
\right] $ on invariant 
$\left( I_{3}^{A_{1}A_{2}A_{3}}\right) _{\left( A_{4}\right)
_{0}}$. The transformed invariant is a combination of five three qubit invariants that is%
\begin{eqnarray}
\left( I_{3}^{A_{1}A_{2}A_{3}}\right) _{\left( A_{4}\right) _{0}}^{\prime }&
=\frac{1}{\left( 1+\left\vert y\right\vert ^{2}\right) ^{2}}\left[ \left(
y^{\ast }\right) ^{4}\left( I_{3}^{A_{1}A_{2}A_{3}}\right) _{\left(
A_{4}\right) _{1}}-4\left( y^{\ast }\right) ^{3}P_{\left( A_{4}\right)
_{1}}^{A_{1}A_{2}A_{3}}\right. \nonumber \\
& \left. +6\left( y^{\ast }\right) ^{2}T_{A_{4}}^{A_{1}A_{2}A_{3}}-4y^{\ast
}P_{\left( A_{4}\right) _{0}}^{A_{1}A_{2}A_{3}}+\left(
I_{3}^{A_{1}A_{2}A_{3}}\right) _{\left( A_{4}\right) _{0}}\right] .
\end{eqnarray}%
Here prime denotes the transformed invariant and additional invariants are
\begin{eqnarray}
T_{A_{4}}^{A_{1}A_{2}A_{3}} &=&\frac{1}{6}\left(
D^{0000}+D^{0001}+D^{0010}+D^{0011}\right) ^{2} \nonumber \\
&&-\frac{2}{3}\left( D_{\left(
A_{3}\right) _{0}}^{000}+D_{\left( A_{3}\right) _{0}}^{001}\right) \left(
D_{\left( A_{3}\right) _{1}}^{000}+D_{\left( A_{3}\right) _{1}}^{001}\right) 
\nonumber \\
&&+\frac{1}{3}\left( D_{_{\left( A_{4}\right) _{0}}}^{000}+D_{_{\left(
A_{4}\right) _{0}}}^{001}\right) \left( D_{_{\left( A_{4}\right)
_{1}}}^{000}+D_{_{\left( A_{4}\right) _{1}}}^{001}\right) \nonumber \\
&&-\frac{2}{3}\left( D_{\left( A_{3}\right) _{0}\left( A_{4}\right)
_{0}}^{00}D_{\left( A_{3}\right) _{1}\left( A_{4}\right)
_{1}}^{00}+D_{\left( A_{3}\right) _{1}\left( A_{4}\right)
_{0}}^{00}D_{\left( A_{3}\right) _{0}\left( A_{4}\right) _{1}}^{00}\right), 
\end{eqnarray}%
 
\begin{eqnarray}
P_{\left( A_{4}\right) _{i_{4}}}^{A_{1}A_{2}A_{3}}& =\frac{1}{2}\left(
D_{_{\left( A_{4}\right) _{i_{4}}}}^{000}+D_{_{\left( A_{4}\right)
_{i_{4}}}}^{001}\right) \left( D^{0000}+D^{0001}+D^{0010}+D^{0011}\right) \nonumber \\
& -\left( D_{\left( A_{3}\right) _{1}\left( A_{4}\right)
_{i_{4}}}^{00}\left( D_{_{\left( A_{3}\right) _{0}}}^{000}+D_{_{\left(
A_{3}\right) _{0}}}^{001}\right) +D_{\left( A_{3}\right) _{0}\left(
A_{4}\right) _{i_{4}}}^{00}\left( D_{_{\left( A_{3}\right)
_{1}}}^{000}+D_{_{\left( A_{3}\right) _{1}}}^{001}\right) \right).
\end{eqnarray}
Five three tangles, constructed from invariants $\left( I_{3}^{A_{1}A_{2}A_{3}}\right) _{\left( A_{4}\right)
_{i_{4}}}$, $\left( I_{3}^{A_{1}A_{2}A_{3}}\right) _{\left( A_{4}\right)
_{i_{4}}}$, $P_{\left( A_{4}\right) _{0}}^{A_{1}A_{2}A_{3}}$, $P_{\left(
A_{4}\right) _{1}}^{A_{1}A_{2}A_{3}}$ , and $T_{A_{4}}^{A_{1}A_{2}A_{3}}$,
capture the entanglement of $A_{1}A_{2}A_{3}$ due to three and four-way
correlations.

\section{Genuine four tangle} Continuing the search for a four qubit 
invariant that detects genuine four-way
correlations, we notice that when a selected U$^{A_{4}}$ results in $%
 \left( I_{3}^{A_{1}A_{2}A_{3}}\right) _{\left( A_{4}\right)
_{0}}^{\prime }=0,$ we have at hand a quartic equation. 
A quartic equation, $y^{4}a-4by^{3}+6y^{2}c-4dy+f=0$, in variable $y$ has
associated polynomial invariants, $S=af-4bd+3c^{2}$, cubic invariant $%
T=acf-ad^{2}-b^{2}f+2bcd-c^{3},$ and discriminant $\Delta =S^{3}-27T^{2}$.
Therefore, the degree eight polynomial invariant associated with $
\left( I_{3}^{A_{1}A_{2}A_{3}}\right) _{\left( A_{4}\right) _{0}}
^{\prime }=0$ is%
\begin{equation}
 I_{(4,8)}^{A_{1}A_{2}A_{3}A_{4}}=3\left(
T_{A_{4}}^{A_{1}A_{2}A_{3}}\right) ^{2}+\left(
I_{3}^{A_{1}A_{2}A_{3}}\right) _{\left( A_{4}\right) _{0}}\left(
I_{3}^{A_{1}A_{2}A_{3}}\right) _{\left( A_{4}\right) _{1}}-4P_{\left(
A_{4}\right) _{0}}^{A_{1}A_{2}A_{3}}P_{\left( A_{4}\right)
_{1}}^{A_{1}A_{2}A_{3}}.
\end{equation}%
The discriminant is given by $\Delta = \left(
I_{4}^{A_{1}A_{2}A_{3}A_{4}}\right)^{3}-27\left(
J^{A_{1}A_{2}A_{3}A_{4}}\right) ^{2}$, where%
\begin{equation}
J^{A_{1}A_{2}A_{3}A_{4}}=\det \left[ 
\begin{array}{ccc}
\left( I_{3}^{A_{1}A_{2}A_{3}}\right) _{\left( A_{4}\right) _{1}} & 
P_{\left( A_{4}\right) _{1}}^{A_{1}A_{2}A_{3}} & T_{A_{4}}^{A_{1}A_{2}A_{3}}
\\ 
P_{\left( A_{4}\right) _{1}}^{A_{1}A_{2}A_{3}} & T_{A_{4}}^{A_{1}A_{2}A_{3}}
& P_{\left( A_{4}\right) _{0}}^{A_{1}A_{2}A_{3}} \\ 
T_{A_{4}}^{A_{1}A_{2}A_{3}} & P_{\left( A_{4}\right) _{0}}^{A_{1}A_{2}A_{3}}
& \left( I_{3}^{A_{1}A_{2}A_{3}}\right) _{\left( A_{4}\right) _{0}}%
\end{array}%
\right] .
\end{equation}%
We may mention here that since there are four ways in which a given set of
three qubits may be selected, $\Delta $ can be expressed in terms of
different sets of three qubit invariants.

The four tangle $\tau_{(4,8)}=4\left\vert \left(  12I_{(4,8)}^{A_{1}A_{2}A_{3}A_{4}}\right)
^{\frac{1}{2}}\right\vert$ quantifies $4-$way correlations \cite{shar131}. If four tangle is zero then
transformation equations acquire a simpler form and yield 
four qubit invariants that quantify $3-$way correlations. Invariant to quantify
entanglement of a four qubit state having purely two qubit correlations can
also be easily obtained. What is the utility of these polynomial invariants?
Quantum entanglement distributed between distinct parties is a physical
resource for practical quantum information processing. Polynomial invariants are used to construct
entanglement monotones to quantify entanglement.

\begin{theacknowledgments}
 Financial support from CNPq Brazil and FAEP UEL Brazil
is acknowledged.
\end{theacknowledgments}

\bibliographystyle{aipproc}

\end{document}